\documentclass[aip,reprint, apl, amsmath, amssymb]{revtex4-1}
\usepackage{graphicx}
\usepackage{dcolumn}
\usepackage{bm}

\begin{document}

\title{Metallic behavior at YBaCuO7/ZAs interfaces (Z=Ga, Al)} 

\author{R. Torres}
  \email{rtorres@fis.cinvestav.mx}   
\author{R. Baquero}
  \email{rbaquero@fis.cinvestav.mx}   
 \affiliation{Centro de Investigaci\'on y de Estudios Avanzados del IPN, Apartado Postal 14-740, M\'exico Distrito Federal 07000, M\'exico.}

\date{\today}

\begin{abstract}
We present the electronic band structure of the interfaces $YBa_2Cu_3O_7/GaAs$ (direct gap) and $YBa_2Cu_3O_7/AlAs$ (indirect gap) in different configurations calculated using the Density Functional Theory as in the Wien2k code within the local density approximation. We have projected the density of states at the atomic layers forming the interface. We concentrated in the semiconductor side. The two first atomic layers in the semiconductor side of the interface present a clear metallic behavior. We found for  both semiconductors considered that it converges towards the bulk atomic-layer projected density of states at the fifth atomic layer from the interface. We considered an ideal non-reconstructed interface in the (001) direction in this work. This behavior is interesting and could be used in several technological applications.
\end{abstract}


\maketitle 

	\section{Introduction}
$YBa_2Cu_3O_7$ is a second generation superconductor with a Tc of the order of  $90K$~\cite{WATHM PRL 87}, Its potential applications include electric hybrid circuits~\cite{BTCBW IEEE 92}, hybrid transistors super- conductor-semiconductor~\cite{F Cry 90}, magnetic microinterruptors~\cite{CGGTB APL 06}, light-emitting diodes~\cite{SKHTA PRL 11} and diode-detectors~\cite{MMS APL 73}, among others~\cite{GKV IEEE 93,GKV patent 95}. The possible change of character at the semiconductor side of a superconductor/semiconductor interface, could allow the semiconductor well establish technology to take advantage of the known properties of superconductors for technological applications.

Interfaces containing YBCO7 have been studied experimentally~\cite{HFB IEEE 91} as well as theoretically~\cite{SCH APL 07} since long ago. Nevertheless, the semiconductor/superconductor interfaces have been scarcely studied except for some oxides~\cite{SIKMN IEEE 91}.

In general, the change of character at an interface is of interest. A metallic behavior, for example, has been obtained at a semiconductor/semiconductor interface~\cite{SCH APL 11}, by introducing vacancies. The metallic behavior at the semiconductor side of a superconductor/semiconductor interface could induce superconductivity at the semiconductor side. This would open the possibility that the nano-circuit technology that has been intensely developed for semiconductors can be employed with the advantages of the superconducting behavior since no heat will be produced. To induce superconductivity in a semiconductor is not new idea. It has been discussed long ago by Marvin Cohen~\cite{C RMP 64,C PR 64}, for example. Also, interfaces with semiconductors as one of the components are of interest in spintronics and spin polarized transport~\cite{SFHZ IEEE 00}.

	\section{Computational Details}
We use in our calculations the Density Functional Theory (DFT)~\cite{HK,KS} with Linearized Augmented Planes Waves plus local orbitals (LAPW+lo)~\cite{LAPW+lo} as implemented in the Wien2k code~\cite{WIEN2k} together with the Local Density Approximation (LDA). In a run $YBa_2Cu_3O_7/GaAs$, we used the Generalized Gradient Approximation (GGA) instead but found no significant differences. We used a cutoff $R_{mt}K_{max}=7.0$ The Muffin-tin radius, $R_{mt}$ was, 2.3 for Y, 2.5 for Ba, 1.79 for Cu, 1.58 for O, and 2.22 for Ga, Al and As in atomic units, when in the configuration show in Fig.~\ref{fig:Estructuras}a. When in the configuration in Fig.~\ref{fig:Estructuras}b and Fig.~\ref{fig:Estructuras}c, $R_{mt}=2.2$ a.u. in the first atomic layer (either $Ga$, $As$ or $Al$) and $R_{mt}=2.0$ a.u. for the secon layer for either $Ga$, $As$ or $Al$. We take YBCO7 in the normal state.

The lattice matched ideal interfaces were built up  with two unit cells of the semiconductor and one of the superconductor which acts as the substrate and therefore the semiconductor takes the lattice parameter of the YBCO7. To avoid tension at the interface as much as possible we have turn the semiconductor unit cell by $45^\circ$ around the z-axis. We considered orthorhombic $YBa_2Cu_3O_7$ (spatial group $Pmmm$), with lattice parameters ~\cite{BSCHJGS APL 87} $a=3.8231$, $b=3.8864$, $c=11.6807$ in \AA. The semiconductor is therefore under tension in the way shown in Table~\ref{tab:PSemi}. The new lattice parameters for the semiconductor ($a_{\Vert}$ and $a_{\bot}$) differ from the original ones ($l_{GaAs}=5.6583$\AA, $l_{AlAs}=5.6611$\AA) and were calculated in the following way. The $x$ and $y$ axis parameters (parallel to the $a$ and $b$ axis of the superconductor) are given by

\begin{equation}
l_{j\Vert}=\sqrt{a^{2}+b^{2}}
\end{equation}

In the $z$-direction, the semiconductor lattice parameter is given by~\cite{B Lib 97}

\begin{equation}
l_{j\bot}=(1-\sigma_{ST}f_j)l_j
\end{equation}

where

\begin{equation}
\frac{\Delta l_j}{l_j}=\frac{l_j-l_{j\Vert}}{l_j}=-f_j.
\end{equation}

The $\sigma^j_{ST}$ parameter is

\begin{equation}
\sigma^j_{ST}=\frac{C^j_{11}}{2C^j_{12}}
\end{equation}

$C^j_{11}$ and $C^j_{12}$ are the elastic constants of the semiconductor.

\begin{table}[h]
\begin{center}
\caption{\label{tab:PSemi}Parameters used in the calculation.}
\begin{tabular}{|l|l|r|l|}
\hline
Semiconductor & $GaAs$ & $AlAs$ \\
\hline
Lattice constant of the bulk $l_j$ (\AA) & 5.6533 & 5.6611 \\
\hline
Parallel lattice constant $l_{j\Vert}$ (\AA) & 5.4516 & 5.4516 \\
\hline
Perpendicular lattice constant $l_{j\bot}$ (\AA) & 5.8760 & 5.8819 \\
\hline
Stress on $l_{j\Vert}$ ($\%$) & 3.57 & 3.70 \\
\hline
Stress on $l_{j\bot}$ ($\%$) & 3.94 & 3.90 \\
\hline
Elastic constant in GPa~\cite{A Lib 07} &   &   \\
$C^j_{11}$ & 118.4 & 125 \\
$C^j_{12}$ & 53.7 & 53.4 \\
\hline
\end{tabular}
\end{center}
\end{table}

	\section{Interfaces Considered}

\begin {figure}[h]
\begin {center}
\includegraphics{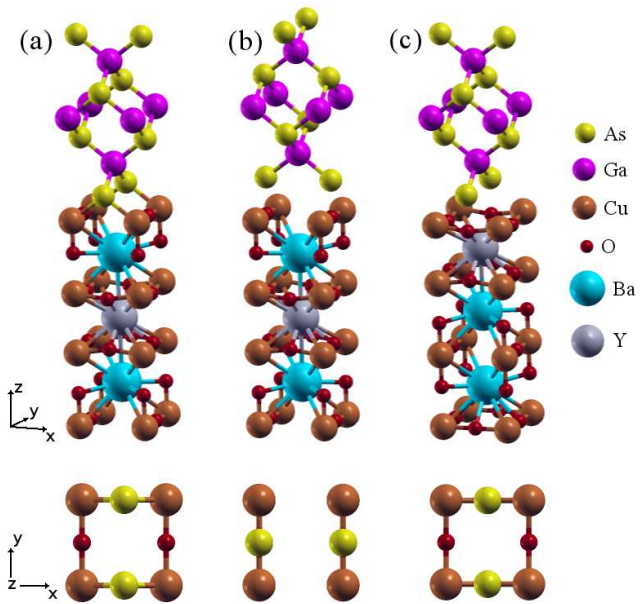}
\end {center}
\vspace*{-3ex}
\caption{(Color online) The interfaces studied a) $chain/X$, b) $chain_O/X$ and c) $plane/X$. See text for the definition of the symbols.}
\label{fig:Estructuras}
\end {figure}

We have performed the calculation of the Interface Density of States (IDOS) for the following systems,

\begin{enumerate}
  \item $CuO$-chains terminated $YBa_2Cu_3O_7$ with both $Ga$ ($As$) terminated $GaAs$ rotated $45^\circ$ clockwise so that the first atomic layer on the semiconductor side faces the sides of $YBa_2Cu_3O_7$ that do not contain oxygen atoms (see Fig.~\ref{fig:Estructuras}a). We label the interfaces accordingly as chain/$X$, $X=AsGa$ (meaning that the $As$ atomic layer is facing $YBa_2Cu_3O_7$),  $chain/GaAs$, $chain/AlAs$ and $chain/AsAl$.
  We took the distance between the adyacent atomic layers ($d$) at the interface as $d=(d_1+d_2)/2$ where $d_1$ and $d_2$ are the distance between the two first atomic layers at each side of the interface.

  \item Same system but now the rotation of the semiconductor is counter clockwise so that the atoms on the first atomic layer of the semiconductor side face the oxygen atoms of the superconductor chains We denote them as $chain_O/X$, where $X$ has the same definition as before (See Fig.~\ref{fig:Estructuras}b).
  We took the distance at the interfaces as $d=2.0\textrm{\AA}$ so that all core states lie inside the muffin-tin sphere.

  \item The third series of calculations were performed with $CuO_2$-plane terminated $YBa_2Cu_3O_7$, which we denote as $plane/X$ where $X$ is defined as before. The rotation was made clockwise. The calculation with the rotation counter clockwise was not performed since the difference in the resulting position of the $O$-atoms is very small (a $0.06\%$ difference with respect to the YBCO7 c-axis length). (See Fig.~\ref{fig:Estructuras}c).
   We took the distance at the interfaces as $d=2.2\textrm{\AA}$ so that all core states lie inside the muffin-tin sphere.
    
\end{enumerate}

	\section{Results and Discussion}

\begin {figure}[bth]
\begin {center}
\includegraphics{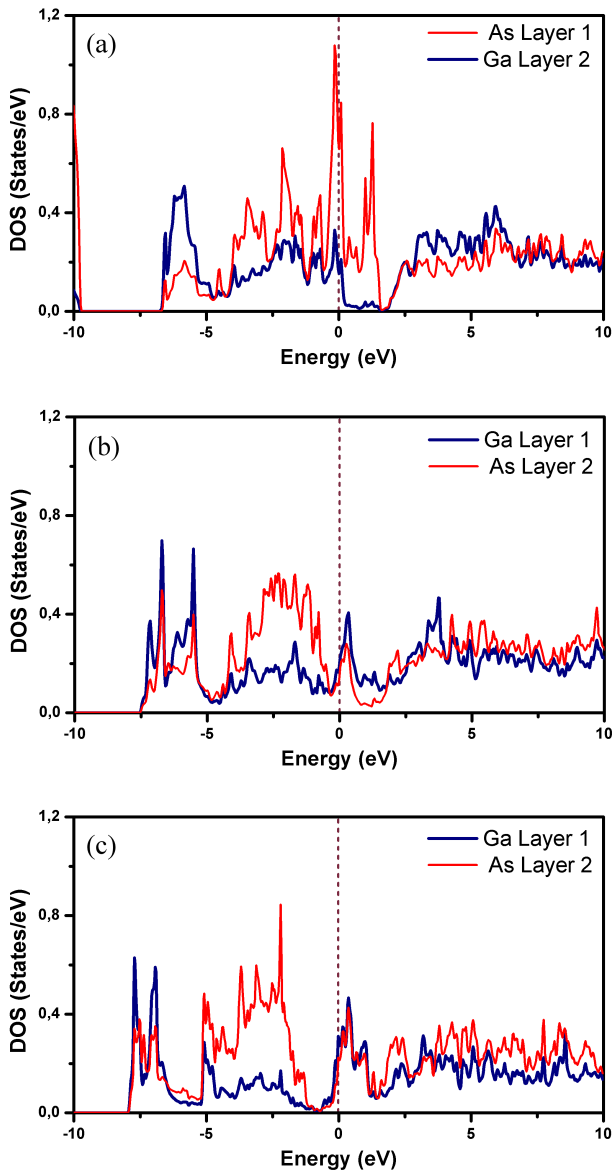}
\end {center}
\vspace*{-3ex}
\caption{(Color online) characteristic results for a)$chain/AsGa$, b)$chain/GaAs$ and c)$plane/ZAs$.}
\label{fig:Resultados1}
\end {figure}

We present in Fig.~\ref{fig:Resultados1}, the Interface Density of States (IDOS) obtained in all the cases considered in this work. A very remarkable result is that in all the configurations considered the first few layers in the semiconductor side present a metallic behavior. This change of character might turn out to be very usefull in technological applications. In more detail, we obtained  three somehow different results presented below. In Fig.~\ref{fig:Resultados1}a, we present the IDOS of the first two atomic layers of the interface $chain/AsGa$. The major contribution in these two layers comes from states of p-symmetry. The Fermi energy is at the origin. As it is apparent from the figure, the two atomic layers from the semiconductor side facing the chains of YBCO7 present a metallic behavior. In Fig.~\ref{fig:Resultados1}b, we present the first two atomic layers of the $chain/GaAs$. Here again we can see that the two atomic layers are metallic.
In Fig.~\ref{fig:Resultados1}c, we present the IDOS for the two first atomic layers on the semiconductor side for $plane/GaAs$ interface. 

In general, our results show that all the interfaces of the form $chain/AsZ$, $chain_O/AsZ$ and $plane/AsZ$ behave as shown in Fig.~\ref{fig:Resultados1}a, the interfaces of the kind $chain/ZAs$, $chain_O/ZAs$ behave as illustrated in Fig.~\ref{fig:Resultados1}b and the interfaces $plane/ZAs$ behave as illustrated in Fig.~\ref{fig:Resultados1}c.

\begin {figure}[bth]
\begin {center}
\includegraphics{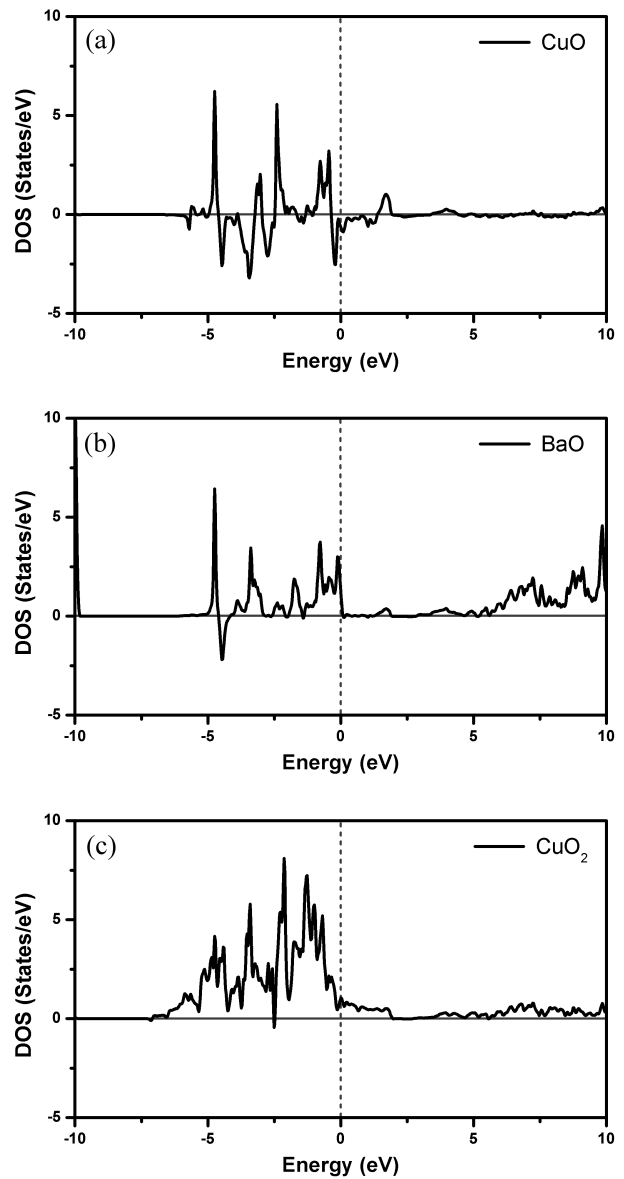}
\end {center}
\vspace*{-3ex}
\caption {Difference between the bulk $YBCO7$ atomic-plane Local Density of States (LDOS) and the IDOS at the same atomic plane for the $chain/AsGa$ case. (a)$CuO$, (b) $BaO$ and (c) $CuO_2$.} 
\label{fig:PlanosDonadores}
\end {figure}

Since at thermal equilibrium the Fermi level is unique for the two sides of the interface, electron states are redistributed. We illustrate this in Fig.~\ref{fig:PlanosDonadores}. The Atomic-Plane Local Density of States (LDOS), is compared to the corresponding Bulk LDOS. We present in Fig.~\ref{fig:PlanosDonadores} the difference (Bulk $-$ Interface) between the two to show explicitly the changes at different frequencies.   

In Fig.~\ref{fig:PlanosDonadores}, we present the difference at the $CuO$-plane (Fig.~\ref{fig:PlanosDonadores}a), the planes $BaO$ (Fig.~\ref{fig:PlanosDonadores}b), and $CuO_2$ (Fig.~\ref{fig:PlanosDonadores}c). The $CuO_2$ plane looses charge the most, as it is evident from Fig.~\ref{fig:PlanosDonadores}c. Looking in more detail, oxygen in  both, the $CuO$ and the $CuO_2$ planes looses about the same amount of charge.  The $Cu$ atoms at these planes loose  the most while the $Ba$ atoms and the $BaO$ planes loose almost nothing and seem not to be most influenced by the interface. This is common to all the different interfaces studies for both, the chain or plane  interfaces.

\begin {figure}[bth]
\begin {center}
\includegraphics{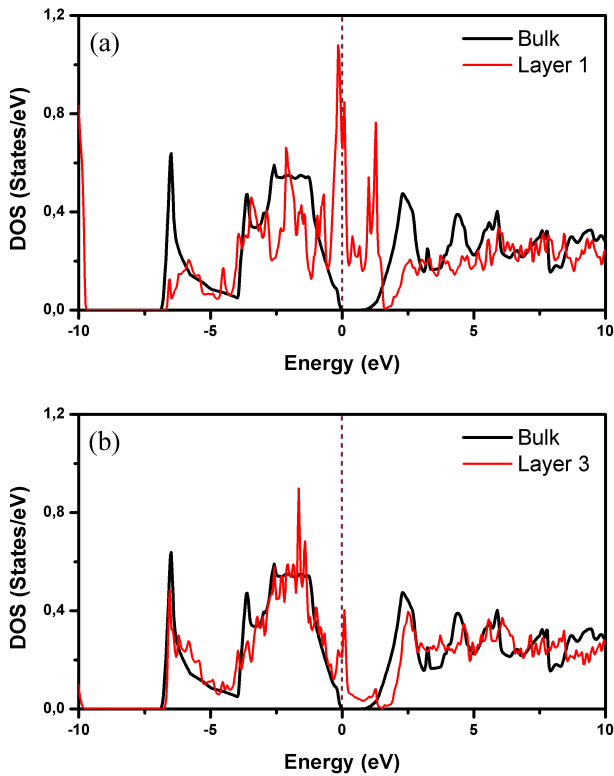}
\end {center}
\vspace*{-3ex}
\caption {(Color online) Here we compare the $As$-atomic layer LDOS when in bulk and at the interface. In (a) right at the interface and (b) on the third layer from the interface.}
\label{fig:resultados2} 
\end {figure}

At the semiconductor side, charge redistribution renders metallic several atomic planes. As the plane distance from the interface enhances, its LDOS recovers the bulk LDOS and the material becomes a semiconductor again.  This is illustrated in Fig.~\ref{fig:resultados2}a.

	\section{Conclusion}
In this work, we have considered  twelve ideal non-reconstructed interfaces (001)$YBCO7/GaZ$ with $Z=Al, As$ and calculated the Local Density of States (LDOS). The interfaces considered, $chain/GaX$, $chain_O/GaX$ and $plane/GaX$, where chain (plane) refers to interfaces where the contact with the semiconductor is with the chain $CuO$ plane ($CuO_2$ plane) of $YBCO7$. The $X$ stands for $Al$ or $As$. The order in which they appear in this work, as $plane/AsGa$, for example, means that the interface begins with an atomic plane of $As$ atoms at the semiconductor side. We distinguish between $chain/GaX$ and $chain_O/GaX$, meaning that in the last case, the $Ga$ atoms face the oxigen atoms of the $CuO$ chains.
The main conclusion is that all the interfaces considered present several planes on the semiconductor side that are metallic. This behavior attenuates slowly and end at the fith layer where the material behaves back as a semiconductor.
This effect is interesting \textit{per se} and it can be also useful for certain technological applications.

\bigskip

 \begin{acknowledgments}
We thank Dr. Amilcar Meneses for useful help in the installation of the WIEN2k code.
 \end{acknowledgments}

\end{document}